\documentclass{acm_proc_article-sp}
\pdfoutput=1

\usepackage{color}
\usepackage{booktabs} 
\usepackage{multirow} 
\usepackage{algorithmic} 
\usepackage{algorithm}
\usepackage{verbatim} 
\usepackage[normalem]{ulem} 
\usepackage{url}  
\usepackage{rotating} 

\definecolor{darkblue}{RGB}{0,0,100}

\usepackage[pdfdisplaydoctitle=true,
colorlinks=true,
citecolor=blue,
linkcolor=black,
urlcolor=darkblue]{hyperref}

\graphicspath{{images/}}

\makeatletter
\let\@copyrightspace\relax
\makeatother

\begin{document}



\hypersetup{pdftitle=Automated Design of Torus Networks}
\hypersetup{pdfauthor=Konstantin S. Solnushkin}
\hypersetup{pdfsubject=Network Design}
\hypersetup{pdfkeywords=Torus network; Data Centre; Supercomputer; Cluster; Automated design}


\title{Automated Design of Torus Networks}

\numberofauthors{1}

\author{
\alignauthor
Konstantin S. Solnushkin\\
       \email{konstantin@solnushkin.org}
}

\maketitle

\begin{abstract}
This paper presents an algorithm to automatically design networks with torus topologies, such as ones widely used in large-scale supercomputers. The characteristic feature of our approach is that real life equipment prices and values of technical characteristics are used. As a result, we also have the opportunity to compare costs of torus and fat-tree networks.

The algorithm is useful as a part of a bigger design procedure that selects optimal hardware of cluster supercomputer as a whole.
\end{abstract}




\section{Introduction}
Torus networks are frequently used in large-scale supercomputers as a cost-efficient alternative to other topologies. Recently it was demonstrated that torus networks for computer clusters can be built from affordable commodity hardware such as InfiniBand.

We describe an algorithm that allows to automatically design torus networks. The algorithm is implemented in a software tool~\cite{solnushkin-network-design-tool-torus}.

This algorithm is intended to be used as a part of a CAD system for cluster supercomputers~\cite{solnushkin2012computer}. Such a system would iterate through different combinations of hardware, varying the number of compute nodes and other parameters. Thus, designing an interconnection network for every hardware combination under review is a self-contained and highly repetitive operation that must be performed efficiently.

During the design process, other characteristics of interconnection networks, such as reliability, can be estimated and used as design constraints or as a part of a complex objective function.

The rest of the article is organized as follows. Section \ref{torus:related-work} describes relevant work in the field of torus networks. Section \ref{torus:gordon} details the design of the ``Gordon'' supercomputer. Section \ref{torus:algorithm} introduces the algorithm, and section \ref{torus:cost-comparison} compares costs of torus networks with fat-trees. Finally, section \ref{torus:conclusions} concludes the article.

\section{Related work}
\label{torus:related-work}

Torus networks have found widespread use in supercomputing. IBM used a 3D torus network in \href{https://en.wikipedia.org/wiki/Blue_Gene}{BlueGene/L}, and a 5D network \cite{chen2011ibm} in BlueGene/Q. A 6D mesh-torus network was used in \href{https://en.wikipedia.org/wiki/K_computer}{``K Computer''} \cite{ajima2012tofu}. Both are \emph{direct networks}, where compute nodes are connected directly to their neighbours, as opposed to \emph{switched fabrics}, where nodes are first connected to switches, and then switches are connected to each other in a torus topology. The example of the latter is a 3D torus network of the \href{http://www.sdsc.edu/supercomputing/gordon}{Gordon} supercomputer \cite{strande2012gordon}.

Torus networks are inherently prone to congestion, but this is mitigated by designers by increasing the number of dimensions. Commenting on the Gordon project, Strande \cite{strande2012torus} quotes the following benefits of torus networks: (a) lower cost compared to fat-trees and (b) easy linear scaling along one of dimensions. However, such scaling may result in unbalanced topologies, leading to bigger latencies and higher congestion on the links in that dimension. Strande also mentions that the torus topology uses short cables, which makes the use of fibre optical cables unnecessary, leading to further cost savings.

Navaridas and Miguel-Alonso \cite{navaridas2011indirect} analysed performance of 2D switch-based torus topologies and fat-trees for up to 7680 compute nodes, on a range of workloads, using simulation techniques. They conclude that performance degradation from using torus networks, compared to fat-trees, can reach 20..40\%, and sometimes more, on communication-intensive workloads, which limits applicability of tori in larger installations.

C\'amara \textit{et al.} \cite{camara2007mixed} introduced the technique to turn unbalanced rectangular 2D and 3D tori to twisted tori by rearranging peripheral links, which improves performance characteristics as well as regains network symmetry.

\section{3D dual-rail torus network of the Gordon supercomputer}
\label{torus:gordon}

Gordon supercomputer \cite{strande2012gordon} uses InfiniBand switches with $P=36$ ports of 4X QDR technology. Switches form a 4x4x4 torus; each switch has 6 neighbours, to which it connects with 3 links, thereby utilizing 18 ports out of 36. 17 more ports are used to connect 16 compute nodes and one I/O node.

The network is \emph{dual-rail}, therefore there are actually two tori made of switches, and compute and I/O nodes have two network interfaces, one of which is used to connect to the switch in the first torus (``rail''), and the other to the second one. Currently, one rail is used for MPI, and the other one for I/O traffic. According to Strande \cite{strande2012torus}, there are plans to use both rails simultaneously to provide failover capabilities and improve bandwidth.

\section{Algorithm for designing torus networks}
\label{torus:algorithm}

We propose the algorithm to calculate the number of switches in a torus network, using as input the number of compute nodes to be interconnected and, optionally, a blocking factor that determines the distribution of ports on a switch between compute nodes and neighbouring switches. The algorithm is suitable to design networks built with commodity hardware, such as Gordon's network.

As torus networks are inherently prone to congestion, imposing additional blocking at the switch level is very disadvantageous. However, sometimes blocking is stipulated by the hardware manufacturer, and cannot be avoided. For example, in \cite{navaridas2011indirect} the hardware under review was a blade chassis equipped with $N=20$ compute nodes and an InfiniBand switch with $P=36$ ports. Only 16 ports of the switch were used to connect it to the outside world, which resulted in $Bl=20/16=1,25$ blocking factor. In order to build torus networks for such hardware with the proposed algorithm, we need to specify the blocking factor as an input.

The algorithm tries to build a network using identical switches with $P_E$ ports. Let us describe the algorithm by stages. In line \ref{alg-torus:trivial-star} we check if the switch has enough ports to connect all $N$ nodes. In this case, we use the star topology with only one switch and exit.

Otherwise, we will build a ring or a torus. In lines \ref{alg-torus:ports-to-nodes}..\ref{alg-torus:resulting-blocking} we calculate the number of switch ports that go to compute nodes and to the neighbouring switches, and then recalculate the blocking factor for the network. On line \ref{alg-torus:calculate-switches} we derive the minimal number of switches required to connect $N$ nodes with a given blocking factor. The actual torus will contain slightly more switches (generally, the increase is within 20\% for small networks, and within several percent for the large ones).

On line \ref{alg-torus:calculate-dimensions}, we use a heuristic to determine the number of torus dimensions, based on the number of switches. It is important to note that there are no hard rules when choosing the number of dimensions. Choosing a low number of dimensions for a high number of compute nodes leads to increased network diameter and therefore latencies. On the other side, choosing a too high number of dimensions for a low number of compute nodes does not provide network performance benefits but results in complex cabling patterns. In the case of direct networks this scenario also requires network adapters with an unnecessarily large number of ports.

The optimal number of dimensions depends on the communication pattern of the application, and can be reliably determined, for any given application, only through benchmarking on real hardware or by using simulation such as in \cite{navaridas2011indirect}. Therefore we relied on using a heuristic.

Currently, the dimension choice heuristic returns the number of dimensions as per Table~\ref{table-torus:dimension-heuristic}, up to $D=5$. The layout of switches in the maximal configuration for that number of dimensions is provided in the last column of the table for reference.

If the heuristic returns $D=1$, then we use the ring topology (line \ref{alg-torus:trivial-ring}). Otherwise, we use the torus topology, and need to calculate the number of switches along each of $D$ dimensions by rounding $\sqrt[D]{E}$ to the nearest integer (line \ref{alg-torus:switches-along-dimensions}).

\renewcommand{\algorithmicrequire}{\textbf{Input:}}
\renewcommand{\algorithmicensure}{\textbf{Goal:}}

\begin{algorithm}
\caption{Design a torus network}
\label{alg-torus}
\begin{algorithmic}[1]
\REQUIRE ~ \\
$N$: Number of nodes to interconnect \\
$Bl$: Blocking factor \\
$P_E$: Number of switch ports
\ENSURE Optimal network structure: \\
$D$: Number of torus dimensions \\
$d=\langle d_1, \dots, d_D \rangle$: Number of switches along each dimension \\
$E$: Total number of switches \\
$Bl_r$: Resulting blocking factor \\
$L$: Number of cables \\
$f$: Objective function for the optimal network structure \\

\IF{$P_E \ge N$}
  \label{alg-torus:trivial-star}
  \STATE \COMMENT{ If there exists a switch with $N$ or more ports }
  \PRINT Topology: star
  \STATE $E \gets 1$; $Bl_r \gets 1$; $L \gets N$
  \STATE Compute $f$
  \STATE Exit
\ENDIF

\STATE $P_{En} \gets \lfloor P_{E} \cdot (Bl / (1 + Bl))\rfloor$ \COMMENT{ Ports to nodes }
\label{alg-torus:ports-to-nodes}
\STATE $P_{Ec} \gets P_{E} - P_{En}$ \COMMENT{ Ports to other switches }
\label{alg-torus:ports-to-core}
\STATE $Bl_r \gets P_{En} / P_{Ec}$ \COMMENT{ Resulting blocking }
\label{alg-torus:resulting-blocking}
\STATE $E \gets \lceil N / P_{En} \rceil$ \COMMENT{ Minimal number of switches }
\label{alg-torus:calculate-switches}
\STATE $D \gets GetDimCount(E)$ \COMMENT { Heuristic for the number of torus dimensions}
\label{alg-torus:calculate-dimensions}

\IF{$D = 1$}
  \PRINT Topology: ring
  \label{alg-torus:trivial-ring}
\ELSE
  \PRINT Topology: torus \\
  \STATE $d_i \gets round(\sqrt[D]{E})~|~i=1 \ldots D-1$ \COMMENT{ Number of switches along dimensions }
  \label{alg-torus:switches-along-dimensions}
  \STATE $d_D \gets \lceil E / d_1^{D-1}\rceil$ \COMMENT{ Switches in the last dimension }
  \label{alg-torus:switches-last-dimension}
  \STATE $E \gets \prod_{i=1}^{D} d_i$ \COMMENT{ Actual number of switches }
  \label{alg-torus:actual-number-of-switches}
\ENDIF
\STATE $L \gets N + E \cdot P_{Ec} / 2$ \COMMENT{ Number of cables }
\label{alg-torus:actual-number-of-cables}
\STATE Compute $f$

\end{algorithmic}
\end{algorithm}

This creates a topology close to an ideal square, cube, etc. Packaging constraints, however, may preclude from using this particular ideal layout, and in the resulting unbalanced torus the number of switches along dimensions may differ significantly. The number of switches, $E$, still remains the same as returned by the algorithm, allowing to correctly calculate equipment cost and other metrics.

\begin{table*}
	\centering
	\begin{tabular}{ l c c c }
		\toprule
		Switch count, $E$ & Topology & Dimensions, $D$ & Max. configuration \\
		\midrule
		2 or 3 & Ring & 1 & --- \\
		\midrule
		up to 36 & \multirow{4}{*}{Torus} & 2 & 6x6 \\
		up to 125 & & 3 & 5x5x5 \\
		up to 2401 & & 4 & 7x7x7x7 \\
		more than 2401 & & 5 & (As appropriate) \\
		\bottomrule
	\end{tabular}
	\caption{Heuristic for the number of torus dimensions}
	\label{table-torus:dimension-heuristic}
\end{table*}

\begin{table*}
	\centering
	\begin{tabular}{ c c c c }
		\toprule
		Compute nodes, $N$ & Dimensions, $D$ & Torus topology & Supercomputer of comparable size \\
		\midrule
		1,000 & 3 & 4x4x4 & Gordon \cite{strande2012gordon}\\
		6,000 & 4 & 4x4x4x6 & Stampede \cite{tacc-stampede}\\
		8,000 & 4 & 5x5x5x4 & Tianhe-1A \cite{nscc-tianhe}\\
		10,000 & 4 & 5x5x5x5 & SuperMUC \cite{lrz-supermuc} \\
		19,000 & 4 & 6x6x6x5 & Titan \cite{titan-ornl} \\
		\bottomrule
	\end{tabular}
	\caption{Sample output for Algorithm~\ref{alg-torus}}
	\label{table-torus:sample-output}
\end{table*}

On the next step, we calculate the number of switches in the last dimension (line \ref{alg-torus:switches-last-dimension}) and recalculate the total number of switches as the product of switch counts along all dimensions (line \ref{alg-torus:actual-number-of-switches}).

The number of cables is determined on line \ref{alg-torus:actual-number-of-cables}. The number of switch ports facing to neighbouring switches, $P_{Ec}$, is divided by two, because two ports are connected with one cable. This is then multiplied by the number of switches $E$. Compute nodes are connected with additional $N$ cables. The network is expandable from $N$ up to $E \cdot P_E$ compute nodes. Inter-switch links run in bundles of approximately $P_{Ec} / (2 \cdot D)$, therefore it is often possible to use cables that integrate several links (such as a 12x InfiniBand cable that integrates three 4x links) to reduce the number of physical cables, simplifying installation.

Sample output of the algorithm for commodity InfiniBand switches with $P_E=36$ ports and a non-blocking network ($Bl=1$) is presented in Table~\ref{table-torus:sample-output}.

\section{Cost comparison of torus and fat-tree networks}
\label{torus:cost-comparison}

We used real life equipment costs provided by Mellanox Technologies to derive costs of fat-tree and torus networks for up to 3,888 compute nodes. We utilized the tool for automated design of cluster interconnection networks \cite{solnushkin-network-design-tool-torus}. Equipment costs are given for the older generation of equipment (InfiniBand QDR), and technical characteristics are summarized in Table~\ref{table-torus:equipment-characteristics}. Cable cost is assumed to be \$80.

We consider three models of switches. The first of them, the 36-port switch, is used for building torus networks, and is also utilized on edge level of fat-tree networks. The other two are modular switches that have 108 and 216 ports in their maximal configurations. The actual number of supported ports depends on the number of installed line cards, which leads to 6 and 12 configurations of these switches, respectively. Each configuration has its own set of technical characteristics as well as cost.

\begin{sidewaystable}
	\centering
	\begin{tabular}{ c c c c c c c }
		\toprule
		Switch applicability & Switch model & Port count & Size, U & Weight, kg & Power, W & Cost, \$ \\[3pt]
		\midrule
		Torus, Fat-tree (edge layer) & Mellanox Grid Director 4036 & 36 & 1 & 7,7 & 202 & 10,820 \\[3pt]
		\cline{1-7} \multirow{6}{*}{Fat-tree (core layer)} & \multirow{6}{*}{Mellanox IS5100} & 18 & \multirow{6}{*}{7} & 75,1 & 516 & 78,500 \\
		& & 36 & & 77,8 & 606 & 90,000 \\
		& & 54 & & 80,6 & 696 & 101,500 \\
		& & 72 & & 83,3 & 786 & 113,000 \\
		& & 90 & & 86,1 & 876 & 124,500 \\
		& & 108 & & 88,9 & 966 & 136,000 \\
		\cline{1-7} \multirow{12}{*}{--''--} & \multirow{12}{*}{Mellanox IS5200} & 18 & \multirow{12}{*}{10} & 115,7 & 516 & 125,500 \\
		& & 36 & & 118,4 & 606 & 137,000 \\
		& & 54 & & 121,2 & 696 & 148,500 \\
		& & 72 & & 123,9 & 786 & 160,000 \\
		& & 90 & & 126,7 & 876 & 171,500 \\
		& & 108 & & 129,5 & 966 & 183,000 \\
		& & 126 & & 132,2 & 1,056 & 194,500 \\
		& & 144 & & 135,0 & 1,146 & 206,000 \\
		& & 162 & & 137,7 & 1,236 & 217,500 \\
		& & 180 & & 140,5 & 1,326 & 229,000 \\
		& & 198 & & 143,3 & 1,416 & 240,500 \\
		& & 216 & & 146,0 & 1,506 & 252,000 \\
		\bottomrule
	\end{tabular}
	\caption{Characteristics of InfiniBand QDR equipment}
	\label{table-torus:equipment-characteristics}
\end{sidewaystable}

\begin{sidewaystable}
	\centering
	\begin{tabular}{ c c c }
		\toprule
		Network type & Non-blocking & 2:1 blocking \\
		\midrule
		Topology & Star & Two-level fat-tree \\
		Edge level switch & Mellanox IS5200 (162 ports) & Mellanox Grid Director 4036 (36 ports) \\
		Core level switch & N/A & Mellanox IS5100 (90 ports) \\
		Power, W & 1,236 & 2,290 \\
		Weight, kg & 137,7 & 140,0 \\
		Size, U & 10 & 14 \\
		Cost, \$ & 229,500 & 218,960 \\
		\bottomrule
	\end{tabular}
	\caption{Structure comparison for two types of fat-tree networks, for $N=150$ nodes.}
	\label{table-torus:structure-comparison}
\end{sidewaystable}

\begin{figure*}
\centering
\includegraphics[width=120mm]{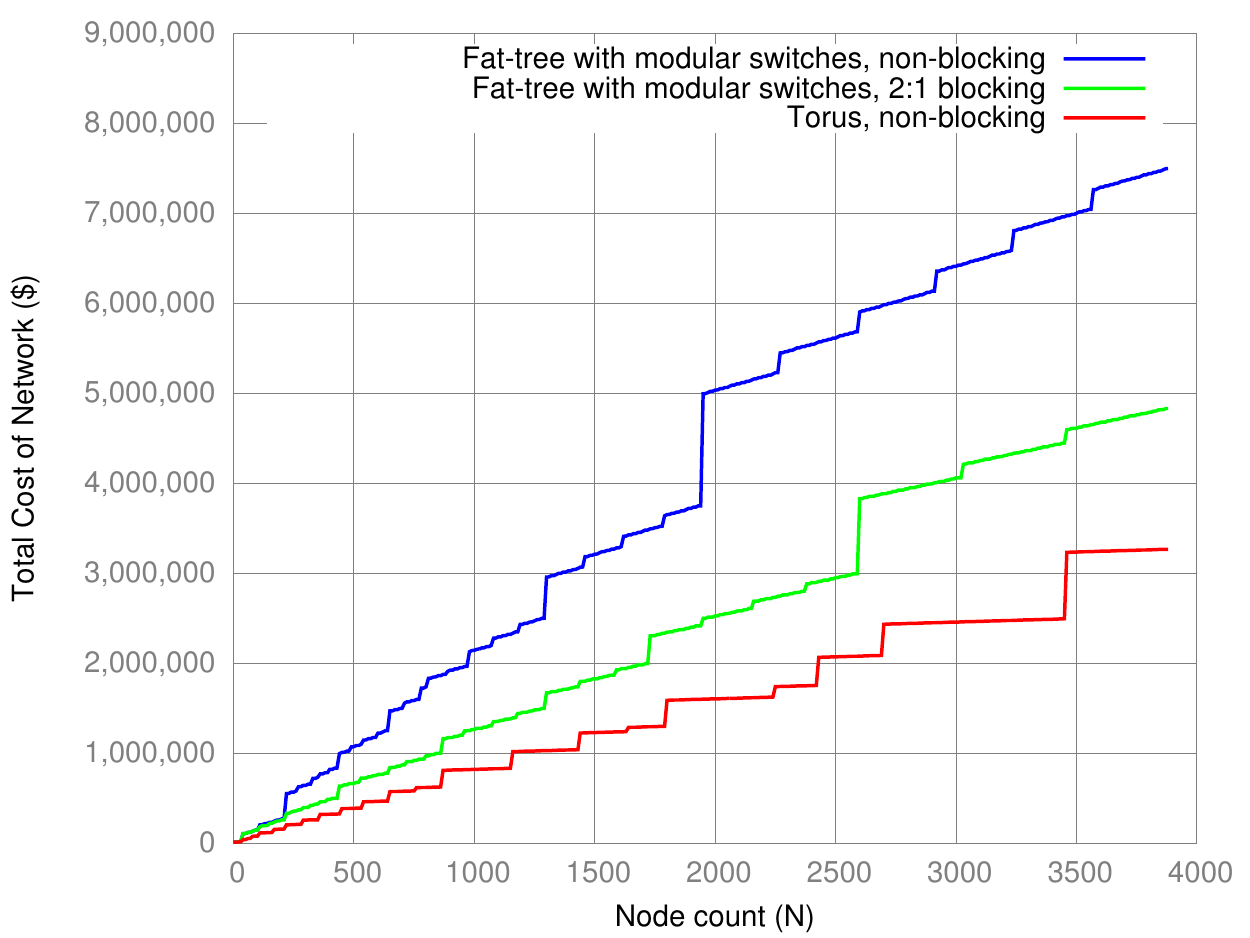}
\caption{Cost comparison of fat-tree and torus networks}
\label{fig-torus:network_cost_comparison}
\end{figure*}

The set of equipment described above allows to build non-blocking fat-tree networks with up to $N_{max}=P_E \cdot P_C/2=36 \cdot 216/2=3888$ nodes. On Fig.~\ref{fig-torus:network_cost_comparison} we plot costs of non-blocking as well as 2:1 blocking fat-tree networks, and torus networks. As expected, the cost of 2:1 blocking fat-trees is lower than of their non-blocking counterparts; but reduction in cost is less than twofold. Torus networks are consistently cheaper than fat-trees; however, their inherent blocking may have detrimental effect on application performance that will not be offset by lower costs.

\begin{figure*}
\centering
\includegraphics[width=120mm]{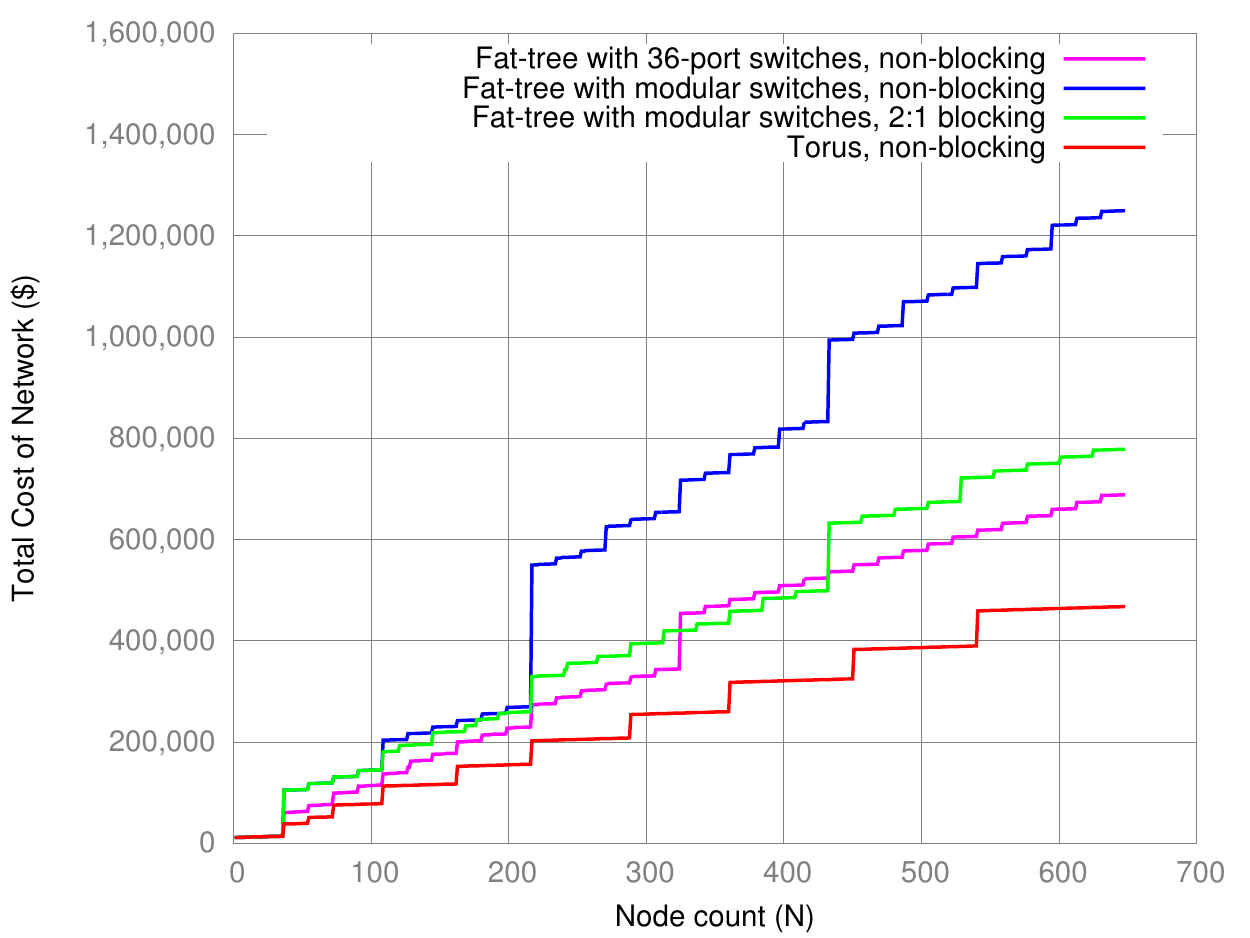}
\caption{Cost comparison of alternative fat-tree building methods}
\label{fig-torus:network_cost_comparison_648}
\end{figure*}

We also consider an alternative way of building fat-trees: using 36-port switches for both core and edge layers. This allows to build non-blocking fat-tree networks with up to $N_{max}=36 \cdot 36/2=648$ nodes. Such networks are characterized by complex wiring patterns between the two layers, but are marginally cheaper to build. Fig.~\ref{fig-torus:network_cost_comparison_648} is essentially a close-up of the previous figure, focusing on values of $N$ up to 648 nodes, with an additional curve representing costs of the alternative fat-tree building method.

As the diagram indicates, using 36-port switches for building fat-trees does indeed lead to certain cost savings: for $N=648$ nodes, per-port cost of such networks is roughly \$1,060, while for the usual way of building fat-trees, using modular switches on the core level, the per-port cost is roughly \$1,930. However, these savings should be weighted against the cost of compute nodes: if the latter is much higher than the per-port cost of the interconnection network, then cost savings might not justify increased wiring and maintenance complexity of this type of networks.

\emph{Example. Let us assume the cost of a compute node is \$5,000. If per-port cost of two types of interconnection networks is \$1,000 and \$2,000, respectively, then savings from using the network of the first type is $7000/6000$, or roughly 17\%. Factoring in costs of other equipment, as well as operating expenses, further dilutes savings.}

Figure~\ref{fig-torus:network_cost_comparison_648} is particularly helpful to emphasize the structure of networks generated by the network design tool \cite{solnushkin-network-design-tool-torus}. Consider, for example, the case of non-blocking and 2:1 blocking fat-trees, for $N=150$ compute nodes. The costs of these two networks are very close, but their structure is entirely different, which is summarized in Table~\ref{table-torus:structure-comparison}.

If the tool is requested to design a non-blocking network, it chooses a star topology with a single modular switch. If, however, a 2:1 blocking network is requested, the result is a two-layer fat-tree, with 36-port switches on the edge level and a 90-port switch on the core level. The latter network is chosen because it is marginally (5\%) cheaper. At the same time, it draws 85\% more power and requires 40\% more space in the rack.

This example illustrates two points: (A) more complex criterion functions, such as total cost of ownership, should preferably be used instead of capital costs; (B) trying to design blocking networks doesn't necessarily save considerable amounts of money, therefore designers should consider non-blocking networks first.

\section{Conclusions}
\label{torus:conclusions}

We presented a simple algorithm for automated design of torus networks. The algorithm relies on a heuristic to choose the number of torus dimensions. We also compared real life costs of torus and fat-tree networks. We found that torus networks are consistently cheaper than non-blocking and 2:1 blocking fat-trees; however, these cost savings may not offset performance penalties, depending on applications used.

%

\bibliographystyle{abbrv}
\bibliography{../text/bibliography/literature}

\begin{thebibliography}{10}

\bibitem{ajima2012tofu}
Y.~Ajima, T.~Inoue, S.~Hiramoto, and T.~Shimizu.
\newblock Tofu: Interconnect for the {K} computer.
\newblock {\em Fujitsu Sci. Tech. J}, 48(3):280--285, 2012.
\newblock \url{http://www.fujitsu.com/downloads/MAG/vol48-3/paper05.pdf}.

\bibitem{camara2007mixed}
J.~C{\'a}mara, M.~Moret{\'o}, E.~Vallejo, R.~Beivide, J.~Miguel-Alonso,
  C.~Mart{\'i}nez, and J.~Navaridas.
\newblock Mixed-radix twisted torus interconnection networks.
\newblock In {\em Parallel and Distributed Processing Symposium, 2007. IPDPS
  2007. IEEE International}, pages 1--10. IEEE, 2007.

\bibitem{chen2011ibm}
D.~Chen, N.~Eisley, P.~Heidelberger, R.~Senger, Y.~Sugawara, S.~Kumar,
  V.~Salapura, D.~Satterfield, B.~Steinmacher-Burow, and J.~Parker.
\newblock The {IBM Blue Gene/Q} interconnection network and message unit.
\newblock In {\em High Performance Computing, Networking, Storage and Analysis
  (SC), 2011 International Conference for}, pages 1--10. IEEE, 2011.
\newblock \url{http://mmc.geofisica.unam.mx/edp/SC11/src/pdf/papers/tp19.pdf}.

\bibitem{lrz-supermuc}
{L}eibniz {R}echenzentrum.
\newblock Super{MUC}.
\newblock \url{http://www.lrz.de/services/compute/supermuc/}.

\bibitem{nscc-tianhe}
{N}ational {S}upercomputing {C}enter in~{T}ianjin ({NSCC}).
\newblock Tianhe-1{A}.
\newblock \url{http://www.nscc-tj.gov.cn/en/}.

\bibitem{navaridas2011indirect}
J.~Navaridas and J.~Miguel-Alonso.
\newblock Indirect cube: A power-efficient topology for compute clusters.
\newblock {\em Optical Switching and Networking}, 8(3):162--170, 2011.

\bibitem{titan-ornl}
{O}ak {R}idge {N}ational~{L}aboratory ({ORNL}).
\newblock Titan.
\newblock \url{http://www.olcf.ornl.gov/titan/}.

\bibitem{solnushkin-network-design-tool-torus}
K.~S. Solnushkin.
\newblock Fat-tree and torus network design tool at {C}luster{D}esign.org.
\newblock \url{http://clusterdesign.org/torus/}.

\bibitem{solnushkin2012computer}
K.~S. Solnushkin.
\newblock Computer cluster design automation using web services.
\newblock In {\em Proceedings of International Supercomputing Conference},
  ISC'12, June 2012.

\bibitem{strande2012torus}
S.~Strande.
\newblock Gordon -- design and performance of a {3D} torus interconnect for
  data intensive computing.
\newblock In {\em Proceedings of HPC Advisory Council Held in Conjunction with
  the International Supercomputing Conference}, 2012.
\newblock
  \url{www.hpcadvisorycouncil.com/events/2012/European-Workshop/Presentations/4_SDSC.pdf}.

\bibitem{strande2012gordon}
S.~Strande, P.~Cicotti, R.~Sinkovits, W.~Young, R.~Wagner, M.~Tatineni,
  E.~Hocks, A.~Snavely, and M.~Norman.
\newblock Gordon: design, performance, and experiences deploying and supporting
  a data intensive supercomputer.
\newblock In {\em Proceedings of the 1st Conference of the Extreme Science and
  Engineering Discovery Environment: Bridging from the eXtreme to the campus
  and beyond}, page~3. ACM, 2012.

\bibitem{tacc-stampede}
{T}exas {A}dvanced {C}omputing~{C}enter ({TACC}).
\newblock Stampede.
\newblock \url{http://www.tacc.utexas.edu/stampede}.

\end{thebibliography}

\end{document}